\documentclass[sigconf,authorversion]{acmart}

\usepackage{booktabs}
\usepackage{makecell}
\usepackage[normalem]{ulem}
\AtBeginDocument{%
  }
\copyrightyear{2026}
\acmYear{2026}
\setcopyright{cc}
\setcctype{by}
\acmConference[SIGIR '26]{Proceedings of the 49th International ACM SIGIR Conference on Research and Development in Information Retrieval}{July 20--24, 2026}{Melbourne, VIC, Australia}
\acmBooktitle{Proceedings of the 49th International ACM SIGIR Conference on Research and Development in Information Retrieval (SIGIR '26), July 20--24, 2026, Melbourne, VIC, Australia}
\acmDOI{10.1145/3805712.3808495}
\acmISBN{979-8-4007-2599-9/2026/07}

\usepackage{fontawesome5}
\newcommand{\githubrepo}{\href{https://github.com/matfu-pixel/GRID}{{\faGithub} matfu-pixel/GRID}}

\title{Gated Bidirectional Linear Attention for Generative Retrieval}

\author{Artem Matveev}
\email{matfu21@yandex.ru}
\orcid{0009-0004-0271-221X}
\affiliation{
  \institution{Yandex}
  \city{Moscow}
  \country{Russia}
}

\author{Vladislav Tytskiy}
\email{vladtytskiy@gmail.com}
\orcid{0009-0003-3960-8689}
\affiliation{
  \institution{Yandex}
  \city{Moscow}
  \country{Russia}
}

\author{Sergei Makeev}
\email{neuralsrg@gmail.com}
\orcid{0009-0003-5451-6475}
\affiliation{
  \institution{Yandex}
  \city{Moscow}
  \country{Russia}
}

\author{Sergei Liamaev}
\email{liamaev.sergei@gmail.com}
\orcid{0009-0009-6316-1091}
\affiliation{
  \institution{Yandex}
  \city{Moscow}
  \country{Russia}
}

\renewcommand{\shortauthors}{Artem Matveev, Vladislav Tytskiy, Sergei Makeev, and Sergei Liamaev}

\begin{abstract}
In recommender systems, generative retrieval typically uses an encoder–decoder setup: an encoder processes a user interaction history, and an autoregressive decoder then generates recommended items. In large-scale streaming services, active users accumulate very long histories over time. As histories grow, the encoder becomes a major latency bottleneck because softmax attention scales quadratically with sequence length. In our experiments, using bidirectional attention in the encoder substantially improves quality. However, most sub-quadratic attention methods focus on causal attention.

We propose Gated Bidirectional Linear Attention (GBLA), a linear-time bidirectional attention layer that extends kernelized linear attention with three lightweight components: local causal mixing (Conv1D), sequence-level key gating for soft forgetting, and a gated RMSNorm output. On a large-scale Yandex Music dataset, a hybrid encoder that interleaves self-attention (SA) and GBLA in a 1:2 ratio (one SA block followed by two GBLA blocks) matches bidirectional self-attention quality. On H100 GPUs, GBLA reaches up to an \(8.2\times\) single-layer speedup at a history length of 32768, compared to FlashAttention-v3. Finally, we show that the same hybrid design generalizes beyond our proprietary setting, consistently preserving self-attention retrieval quality on public Amazon benchmarks.
\end{abstract}



\begin{document}
\maketitle

\section{Introduction}
Sequential modeling in recommendation learns from an ordered user interaction history to predict future actions. Transformer models are widely used for sequential modeling in large-scale recommendation systems at Google\,\cite{rajput2023tiger}, Pinterest\,\cite{pancha2022pinnerformer}, Yandex\,\cite{khrylchenko2025argus}, ByteDance\,\cite{chai2025longer}, and others. In particular, generative retrieval models often adopt an encoder-decoder architecture, where the encoder processes a user interaction sequence and the decoder generates recommended items. Due to their efficiency in end-to-end recommendation generation, such generative retrieval models have been deployed at major streaming services, including YouTube\,\cite{singh2024better} and KuaiShou\,\cite{deng2025onerec}. 

In sequential modeling, the user interaction sequence is the input to the transformer, and increasing the number of interactions is known to improve recommendation quality\,\cite{zhai2024actions, khrylchenko2025argus}. However, transformer architectures rely on self-attention, whose time and memory cost scales quadratically with sequence length. As histories grow, this quadratic complexity can make latency prohibitive for online services. Existing solutions either compress the input sequence\,\cite{chai2025longer}, or rely on systems-level optimizations such as context parallelism\,\cite{dong2025contextparallelism} and hierarchical history pathways\,\cite{zhou2025onerec_tr}. Albeit effective, sequence compression can discard information that is important for retrieval quality, and systems-level methods do not remove the quadratic dependence of attention on sequence length.

In large language models, where input lengths can reach millions of tokens, the quadratic cost of attention makes training and inference very expensive. To address this limitation, \citet{katharopoulos2020linear} proposed kernelized linear attention, which rewrites attention as a dot-product in a kernel feature space and reduces complexity from \(\mathcal{O}(n^2)\) to \(\mathcal{O}(n)\). Subsequent work, such as Mamba\,\cite{gu2023mamba} and delta-rule variants with improved hardware efficiency\,\cite{yang2024gateddeltanet}, further advanced long-context modeling. However, most of these methods focus on causal attention. In generative retrieval, by contrast, the encoder accounts for most of the computational cost and empirically benefits from bidirectional masking\,\cite{rajput2023tiger}. Scalable bidirectional linear attention is therefore important but remains underexplored. Existing bidirectional linear attention frameworks, such as LION\,\cite{afzal2025lion}, achieve linear complexity only at inference time, while training still relies on quadratic attention.

We introduce Gated Bidirectional Linear Attention (GBLA), a linear-time bidirectional attention mechanism designed for long-context generative retrieval. GBLA reduces training and inference latency relative to optimized quadratic attention implementations such as FlashAttention-v3\,\cite{shah2024flashattention3}, while maintaining near-parity retrieval quality. We evaluate GBLA on both in-house and public benchmarks, and release our code at \githubrepo.

\section{Method}
We study generative retrieval with an encoder-decoder transformer in a setting similar to OneRec\,\cite{deng2025onerec}. Given a user-item interaction sequence \(x^u = (i^u_{1}, i^u_{2}, \ldots, i^u_{L})\), generative retrieval aims to generate a sequence of items for user \(u\) to interact with next. In many services, the recommendation system returns ranked items in batches; as the user consumes one batch, the system serves the next, naturally splitting sequence of interactions into batches. We treat the interactions from a single batch as target items and train the model to generate them given the sequence of previous interactions.

To embed item \(i^u_k\), we use the multi-hash technique\,\cite{tito2017hash}: the item ID is mapped by different hash functions to several entries in a shared embedding table, the obtained embeddings are concatenated, and the resulting vector is projected with a linear layer. The sequence of embedded items is then passed through a bidirectional transformer encoder. The decoder conditions on the output of the encoder using cross-attention and autoregressively generates semantic IDs\,\cite{rajput2023tiger} of candidate items.

\paragraph{Bidirectional softmax attention.} 
Let \(\mathbf{X}\in\mathbb{R}^{L\times d}\) be an embedded sequence of user interactions. For a single attention head with head dimension \(d_h\), let \(\mathbf{Q} = \mathbf{X}\mathbf{W}_Q\), \(\mathbf{K} = \mathbf{X}\mathbf{W}_K\), \(\mathbf{V} = \mathbf{X}\mathbf{W}_V\), where \(\mathbf{W}_Q, \mathbf{W}_K, \mathbf{W}_V \in \mathbb{R}^{d \times d_h}\). Standard bidirectional self-attention has quadratic complexity in the input sequence length \(L\) and for a single query \(\mathbf{q}_i\) is given by
\[
\mathrm{Attn}(\mathbf{q}_i, \mathbf{K}, \mathbf{V})  = \frac{\sum_{j=1}^L \exp\left(\mathbf{q}_i^\top \mathbf{k}_j / \sqrt{d_h}\right)\mathbf{v}_j^\top}{\sum_{j=1}^L\exp\left(\mathbf{q}_i^\top \mathbf{k}_j / \sqrt{d_h}\right)}.
\]

\textit{Bidirectional Linear Attention.} In linear attention\,\cite{katharopoulos2020transformers}, exponential similarity score \(\mathrm{sim}(\mathbf{q},\mathbf{k}) = \exp(\mathbf{q}^\top \mathbf{k} / \sqrt{d_h})\) is replaced by dot product in a kernel feature space: \(\mathrm{sim}(\mathbf{q},\mathbf{k}) = \phi(\mathbf{q})^\top \phi(\mathbf{k})\), where \(\phi\) is an element-wise non-negative feature map, e.g., \(\phi(x) = \mathrm{elu}(x) + 1\). Namely,
\[
\mathrm{Attn_{BLA}}(\mathbf{q}_i, \mathbf{K}, \mathbf{V}) = \frac{\phi(\mathbf{q}_i)^\top\sum_{j=1}^L \phi(\mathbf{k}_j)\mathbf{v}_j^\top}{\phi(\mathbf{q}_i)^\top\sum_{j=1}^L\phi(\mathbf{k}_j)},
\]
and in matrix form
\begin{equation}
    \mathrm{Attn_{BLA}}(\mathbf{Q},\mathbf{K},\mathbf{V}) = \bigl(\phi(\mathbf{Q}) (\phi(\mathbf{K})^\top \mathbf{V})\bigr) \oslash \bigl(\phi(\mathbf{Q}) (\phi(\mathbf{K})^\top \mathbf{1})\bigr).
    \label{eq:bla}
\end{equation}
Here \(\mathbf{1} \in \mathbb{R}^{L}\) is an all-ones vector, and \(\oslash\) denotes element-wise division with broadcasting.
This formula avoids explicit \(L \times L\) attention matrix and reduces complexity from \(\mathcal{O}(L^2 d_h)\) to \(\mathcal{O}(L d_h^2)\).

\begin{figure}[t]
    \centering
    \includegraphics[width=1\linewidth]{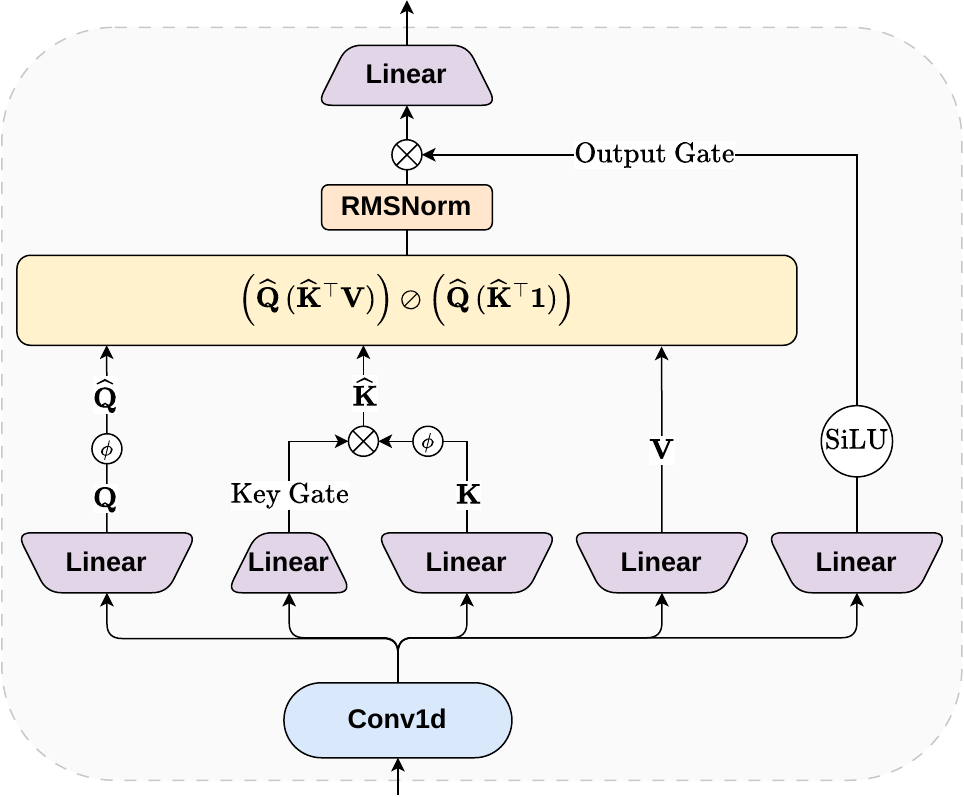}
    \caption{Gated Bidirectional Linear Attention architecture.}
    \label{fig:gbla_achitecture}
\end{figure}

\subsection{Gated Bidirectional Linear Attention.} We introduce Gated Bidirectional Linear Attention (GBLA) (\autoref{fig:gbla_achitecture}), which augments \autoref{eq:bla} with three inexpensive components while preserving linear complexity:
\begin{enumerate}
    \item \textbf{Conv1D.} We apply a 1D convolution on input embeddings to improve short-range pattern extraction. We compute QKV-projections after the convolution:
    \[
    \widetilde{\mathbf{X}} = \mathrm{Conv1D}(\mathbf{X}), \quad \mathbf{Q} = \widetilde{\mathbf{X}}\mathbf{W}_Q,~\mathbf{K} = \widetilde{\mathbf{X}}\mathbf{W}_K,~\mathbf{V} = \widetilde{\mathbf{X}}\mathbf{W}_V.
    \]
    \item \textbf{Key gating.} We learn a per-token scalar gate for keys, which acts as a soft forgetting mechanism, weighting history events according to their importance:
    \[
    \mathrm{Attn_{GBLA}}(\mathbf{q}_i, \mathbf{K}, \mathbf{V}) = \frac{\phi(\mathbf{q}_i)^\top\sum_{j=1}^L g_j \phi(\mathbf{k}_j)\mathbf{v}_j^\top}{\phi(\mathbf{q}_i)^\top\sum_{j=1}^L g_j \phi(\mathbf{k}_j)},
    \]
    where \(g = \mathrm{softmax}(\widetilde{\mathbf{X}}\mathbf{w}_g)\) and \(\mathbf{w}_g\in\mathbb{R}^{d}\). In matrix form:
    \[
    \mathrm{Attn_{GBLA}}(\mathbf{Q},\mathbf{K},\mathbf{V}) = \bigl(\widehat{\mathbf{Q}}\, \bigl(\widehat{\mathbf{K}}^\top \mathbf{V}\bigr)\bigr) \oslash \bigl(\widehat{\mathbf{Q}}\, \bigl(\widehat{\mathbf{K}}^\top\mathbf{1}\bigr)\bigr),
    \]
    where we adopt the notation \(\widehat{\mathbf{Q}} = \phi(\mathbf{Q}),~\widehat{\mathbf{K}} = g \odot \phi(\mathbf{K})\), and \(\odot\) denotes element-wise product with broadcasting.
    \item \textbf{RMSNorm with gating.} Following common long-context designs\,\cite{yang2024gateddeltanet}, we apply gated normalization to the attention output \(\mathbf{O} = \mathrm{Attn_{GBLA}}(\mathbf{Q},\mathbf{K},\mathbf{V}) \in \mathbb{R}^{L \times d_h}\):
    \[
    \mathrm{RmsNormGated}(\mathbf{O}) = \mathrm{RmsNorm}(\mathbf{O}) \odot \mathrm{SiLU}(\widetilde{\mathbf{X}} \mathbf{W}_R),
    \]
    where \(\mathbf{W}_R \in \mathbb{R}^{d \times d_h}\).
\end{enumerate}

As in standard attention, we use multiple attention heads. For each head \(h\), separate matrices \(\mathbf{W}_Q^{h}\), \(\mathbf{W}_K^{h}\), \(\mathbf{W}_V^{h}\), \(\mathbf{w}_g^{h}\), \(\mathbf{W}_R^{h}\) are used to obtain head-specific projections.

\section{Experiments}

In practice, replacing self-attention (SA) with linear attention (LA) is rarely optimal\,\cite{kimiteam2025kimilinear}. We therefore use a hybrid encoder that interleaves SA and LA in \(1:2\) pattern with \texttt{[SA, LA, LA]} ordering. For BLA, we use a fused Triton kernel, while GBLA is optimized with \texttt{torch.compile}. We run all experiments on NVIDIA H100 GPUs rented via a third-party compute provider. 

\subsection{Industrial Setup}

We compare the self-attention encoder with the hybrid encoder on a large-scale production dataset from Yandex Music streaming platform. Training uses 7 consecutive days of interactions, and the training data is consumed in chronological order. Evaluation uses the following day. The training set contains a subsample of 400M training samples.

We find that our setting benefits from shifting most model capacity to the encoder. Specifically, the encoder has \(9\) bidirectional transformer layers with hidden size \(d{=}1024\) and \(16\) attention heads, a subset of which can be replaced with GBLA layers, while the decoder is a single causal layer with both self-attention and cross-attention. For GBLA we set the Conv1D kernel size to \(4\). Since the encoder dominates the computation, we incorporate linear attention into the encoder.

We use a linear learning rate schedule with warmup: the learning rate increases from \(10^{-5}\) to \(x\) over the first \(3000\) iterations and is then linearly decayed to \(0.1x\). Notably, the optimal peak learning rate depends on the architecture: for the full self-attention, we use \(x{=}7{\times}10^{-4}\), while for the hybrid encoder, we use \(x{=}5{\times}10^{-4}\). All experiments use the same effective batch size.

We report retrieval quality as Recall@\( \{10,100,1000\} \). Unless stated otherwise, we fix the maximum history length to \(L=2048\) for the main ablations and separately study scaling to longer histories.

\subsection{Results}

We focus on Recall@\(1000\) since our production pipeline performs retrieval and forwards the top-\(1000\) candidates to the downstream ranker. In this regime, even a \(1\%\) absolute drop in Recall@\(1000\) is significant.

\begin{table}[t]
  \centering
  \caption{Comparison of causal and bidirectional encoder attention mask at length 2048.}
  \label{tab:causal_vs_bidirectional}
  \Description{Comparison of causal and bidirectional mask at length 2048.}
  \small
  \setlength{\tabcolsep}{6pt}
  \begin{tabular}{lccc}
    \toprule
    Model & Recall@10 & Recall@100 & Recall@1000 \\
    \midrule
    Bidirectional SA mask
      & 0.2800 & 0.6150 & 0.8667 \\
    Causal SA mask
      & 0.1878 & 0.5130 & 0.8353 \\
    \bottomrule
  \end{tabular}
\end{table}

\paragraph{Bidirectionality is critical.}
\autoref{tab:causal_vs_bidirectional} highlights the importance of using a bidirectional attention mask in the encoder for generative retrieval.

\begin{table}[t]
  \centering
  \caption{Comparison of hybrid and fully GBLA architecture at length 2048.}
  \label{tab:hybrid_vs_linear}
  \Description{Comparison of hybrid and fully GBLA architecture at length 2048.}
  \small
  \setlength{\tabcolsep}{6pt}
  \begin{tabular}{lccc}
    \toprule
    Model & Recall@10 & Recall@100 & Recall@1000 \\
    \midrule
    Hybrid \texttt{[SA, LA, LA]}
      & 0.2780 & 0.6143 & 0.8668 \\
    Fully GBLA
      & 0.2607 & 0.5940 & 0.8586 \\
    \bottomrule
  \end{tabular}
\end{table}

\paragraph{Hybrid encoder helps.} \autoref{tab:hybrid_vs_linear} shows that fully removing self-attention is typically suboptimal; our best models use a hybrid stack with a \(2{:}1\) ratio of linear attention to self-attention and ordering \texttt{[SA, LA, LA]}.

\begin{table}[t]
  \centering
  \caption{Comparison of bidirectional self-attention and linear attention for different sequence lengths.}
  \label{tab:seqlen_bla_bsa}
  \Description{Comparison of bidirectional self-attention and linear attention for different sequence lengths.}
  \small
  \setlength{\tabcolsep}{6pt}
  \begin{tabular}{c l ccc}
    \toprule
    SeqLen & Method & Recall@10 & Recall@100 & Recall@1000 \\
    \midrule
     512
      & Bidirectional SA & 0.2089 & 0.5072 & 0.8182 \\
      & Hybrid GBLA      & \textbf{0.2102} & \textbf{0.5099} & \textbf{0.8198} \\
    \midrule
    2048
      & Bidirectional SA & \textbf{0.2800} & \textbf{0.6150} & 0.8667 \\
      & Hybrid GBLA      & 0.2780 & 0.6143 & \textbf{0.8668} \\
    \midrule
    4096
      & Bidirectional SA & \textbf{0.2935} & \textbf{0.6385} & 0.8784 \\
      & Hybrid GBLA      & 0.2897 & 0.6368 & \textbf{0.8790} \\
    \midrule
    8192
      & Bidirectional SA & \textbf{0.3013} & \textbf{0.6517} & \textbf{0.8854} \\
      & Hybrid GBLA      & 0.2948 & 0.6462 & 0.8842 \\
    \bottomrule
  \end{tabular}
\end{table}

\paragraph{Quality preservation with increasing sequence length.} \autoref{tab:seqlen_bla_bsa} shows that, as measured by Recall@\(1000\), GBLA performs slightly better at sequence lengths \(L \in \{512, 2048, 4096\}\), while at \(L{=}8192\) it incurs only a tiny drop.

\subsection{Ablation}

\begin{table}[t]
  \centering
  \caption{Ablations at history length 2048.}
  \label{tab:industrial_recall_2048_ablation}
  \Description{Ablations at history length 2048}
  \small
  \setlength{\tabcolsep}{6pt}
  \begin{tabular}{lccc}
    \toprule
    Model & Recall@10 & Recall@100 & Recall@1000 \\
    \midrule
    \textbf{GBLA}
      & \textbf{0.2780} & \textbf{0.6143} & \textbf{0.8668} \\
    W/o Key gating
      & \underline{0.2776} & \underline{0.6131} & \underline{0.8661} \\
    W/o Conv1D
      & 0.2747 & 0.6110 & 0.8652 \\
    W/o Gated RMSNorm
      & 0.2726 & 0.6080 & 0.8641 \\
    W/o all (=BLA)
      & 0.2675 & 0.6021 & 0.8620 \\
    \bottomrule
  \end{tabular}
\end{table}

\autoref{tab:industrial_recall_2048_ablation} presents the ablation of individual GBLA components. Each component contributes incrementally to quality: removing Key gating, Conv1D, or gated RMSNorm slightly degrades recall, while removing all add-ons produces the largest drop, especially in Recall@10.

\subsection{Performance Benchmarks}

\begin{figure}[t]
    \centering
    \includegraphics[width=1\linewidth]{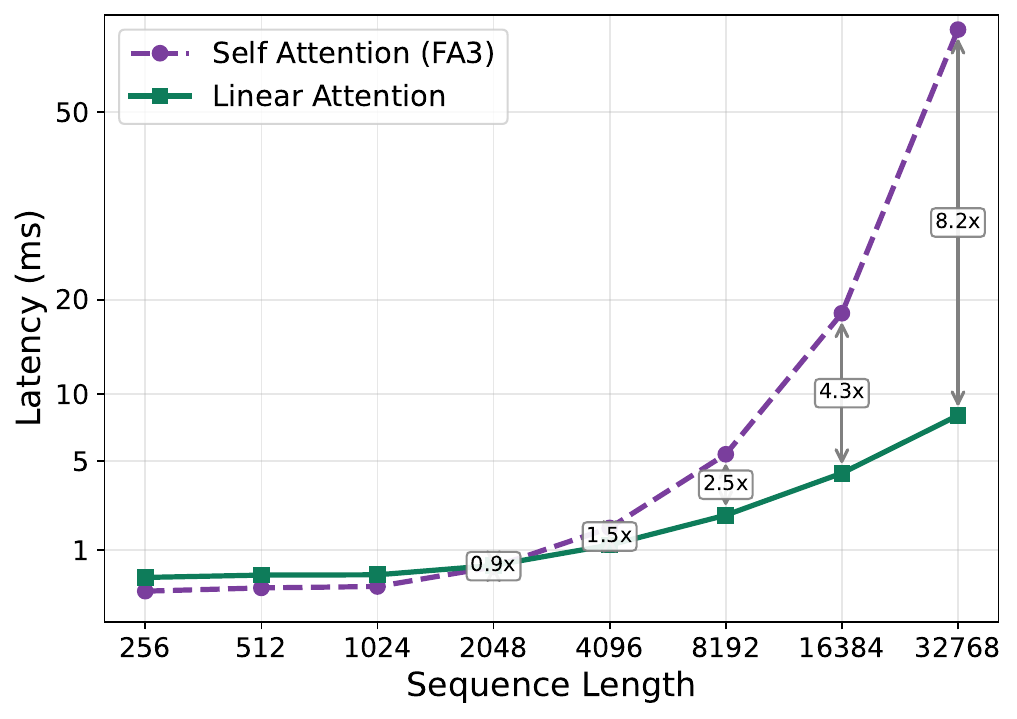}
    \caption{Encoder latency for different sequence lengths.}
    \label{fig:attention_speed}
\end{figure}

\begin{table}[t]
  \centering
  \caption{Per-layer latency on NVIDIA H100 (batch=8).}
  \label{tab:attn_latency_2048_plus}
  \Description{Per-layer latency on NVIDIA H100 (batch=8).}
  \small
  \setlength{\tabcolsep}{6pt}
  \begin{tabular}{rccc}
    \toprule
    SeqLen & Self-Attn (ms) & GBLA (ms) & Speedup (SA/GBLA) \\
    \midrule
         2048  & 0.585  & 0.618  & 0.95\( \times \) \\
     4096  & 1.706  & 1.141  & 1.50\( \times \) \\
     8192  & 5.419  & 2.199  & 2.46\( \times \) \\
    16384  & 18.353 & 4.266  & 4.30\( \times \) \\
    32768  & 67.542 & 8.217  & 8.22\( \times \) \\
    \bottomrule
  \end{tabular}
\end{table}

\paragraph{Inference latency of a single layer.} We compare inference latency of a single GBLA layer against FlashAttention-v3 on H100 GPUs with batch size \(8\). We report the average latency over \(10\) runs. \autoref{tab:attn_latency_2048_plus} shows that GBLA becomes faster starting at \(L{=}4096\), and the speedup grows to \(8.2\times\) at \(L{=}32768\). Importantly, for shorter histories (\(L \le 2048\)), self-attention accounts for substantially less than \(25\%\) of the end-to-end inference time, so the small per-layer slowdown has a negligible impact on overall latency. As \(L\) grows, the self-attention blocks become the dominant bottleneck, making it critical to accelerate attention specifically in the long-context regime. See \autoref{fig:attention_speed} for a more fine-grained comparison. 

\begin{table}[t]
  \centering
  \caption{Training step time on NVIDIA H100: hybrid vs.\ fully self-attention (batch=32 for L=2048, batch=8 for L=8192).}
  \label{tab:train_latency}
  \Description{Training step time on NVIDIA H100 comparing hybrid models to a fully self-attention baseline across sequence lengths.}
  \small
  \setlength{\tabcolsep}{6pt}
  \begin{tabular}{lcccc}
    \toprule
    Method & $L$=2048 (ms) & Speedup & $L$=8192 (ms) & Speedup \\
    \midrule
    SA                   & 3000 & 1.00$\times$ & 5400 & 1.00$\times$ \\
    BLA                  & 2800 & 1.07$\times$ & 3800 & 1.42$\times$ \\
    GBLA                 & 3100 & 0.95$\times$ & 4200 & 1.29$\times$ \\
    \bottomrule
  \end{tabular}
\end{table}

\paragraph{Training speedup of the full hybrid model.} \autoref{tab:train_latency} reports end-to-end training step time on H100 GPUs, averaged over \(10\) runs. Replacing a part of the encoder self-attention blocks with a plain bidirectional linear-attention (i.e. removing all GBLA add-ons such as key gating, Conv1D mixing and gated RMSNorm) reduces step time from \(3000\)\,ms to \(2800\)\,ms at \(L{=}2048\) (\(1.07\times\) speedup) and batch size \(32\), and from \(5400\)\,ms to \(3800\)\,ms at \(L{=}8192\) (\(1.42\times\) speedup) and batch size \(8\). The GBLA hybrid variant shows only a negligible slowdown at \(L{=}2048\), since at this length the FlashAttention-based self-attention blocks account for a small fraction of end-to-end step time.

\subsection{Academic Setup.}

To benchmark GBLA on public data, we adopt the experimental protocol of the GRID framework\,\cite{ju2025grid}. We use our fork of the official implementation, extended with support for evaluating linear-attention variants, available at \githubrepo.

Following GRID, we evaluate on the 5-core filtered Amazon \textsc{Beauty}, \textsc{Sports}, and \textsc{Toys} datasets, using the last item of each user sequence for test, the second-to-last for validation, and the remaining interactions for training. We use the same item text fields (Title, Categories, Description, Price) and extract semantic item embeddings by mean pooling the final hidden states of Flan-T5 (\texttt{Large/XL/XXL}). For semantic ID tokenization, we follow GRID and consider RK-Means, using the same tokenizer training recipe (8 GPUs, per-device batch size 2048).

GRID implements Tiger\,\cite{rajput2023tiger} generative model. For that model, we match GRID's training setup (Adam, learning rate \(5{\times}10^{-4}\), weight decay \(10^{-6}\), batch size 256) and data sampling/selection strategy (sliding-window sampling and early stopping based on validation NDCG@10). We keep the GRID backbone (8 transformer layers total for encoder-decoder variants, 6 attention heads, embedding size 128, and MLP hidden size 1024) and change only the encoder attention operator: we replace a subset of encoder self-attention layers with GBLA, following our hybrid architecture design. We set the Conv1D kernel size to \(4\). We report Recall@K and NDCG@K for \(K\in\{5,10\}\) on the test set, selecting the checkpoint with the best validation Recall@10, and average results over 5 runs, consistent with GRID.

\begin{table}[t]
  \centering
  \caption{Comparison of Tiger and Tiger+GBLA on Amazon datasets, averaged over 5 runs with different seeds.}
  \label{tab:gr_tiger_gbla}
  \Description{Comparison of Tiger and Tiger+GBLA on Amazon datasets, averaged over 5 runs with different seeds.}
  \small
  \setlength{\tabcolsep}{6pt}
  \begin{tabular}{lcccc}
    \toprule
    Model & Recall@5 & Recall@10 & NDCG@5 & NDCG@10 \\
    \midrule

    \multicolumn{5}{c}{\textbf{Beauty}} \\
    \midrule
    Tiger      & \textbf{0.0439} & \textbf{0.0641} & \textbf{0.0289} & \textbf{0.0355} \\
    Tiger+GBLA & 0.0410          & 0.0611          & 0.0273          & 0.0338 \\
    \midrule

    \multicolumn{5}{c}{\textbf{Toys}} \\
    \midrule
    Tiger      & \textbf{0.0402} & \textbf{0.0584} & \textbf{0.0274} & \textbf{0.0333} \\
    Tiger+GBLA & 0.0384          & 0.0579          & 0.0249          & 0.0311 \\
    \midrule

    \multicolumn{5}{c}{\textbf{Sports}} \\
    \midrule
    Tiger      & \textbf{0.0229} & \textbf{0.0345} & \textbf{0.0150} & \textbf{0.0188} \\
    Tiger+GBLA & 0.0218          & 0.0329          & 0.0144          & 0.0180 \\
    \bottomrule
  \end{tabular}
\end{table}

\autoref{tab:gr_tiger_gbla} shows that Tiger and Tiger+GBLA achieve very similar results across all three Amazon datasets. Consistent with the variability observed in GRID’s reported results and ablations, these small deltas are typically not considered significant. Since the user histories are short (at most 128 items), self-attention does not correspond to a major bottleneck. We therefore do not benchmark the performance efficiency in the academic setup.

\section{Conclusion}

We introduced Gated Bidirectional Linear Attention (GBLA), a linear-time bidirectional attention layer for long-history generative retrieval. In our experiments, a hybrid encoder that interleaves self-attention and GBLA matches the bidirectional self-attention quality on a large-scale Yandex Music dataset and preserves similar retrieval metrics on public Amazon benchmarks. We also show that bidirectional masking is crucial: switching the encoder to a causal mask causes a large drop in recall.

GBLA is most useful for long sequences. On H100 GPUs, it becomes faster than FlashAttention-v3 starting at length 4096, and the speedup grows with history length, reaching up to \(8.2\times\) per-layer speedup at length 32768. Overall, GBLA provides a practical way to scale generative retrieval to longer user histories without sacrificing retrieval quality.

\section*{Author Bio}
Artem Matveev is an MSc student in Applied Mathematics and Computer Science at Higher School of Economics and a senior deep learning engineer at Yandex, working on generative recommendation approaches for large-scale recommender systems.

\bibliographystyle{ACM-Reference-Format}
\bibliography{refs}

\end{document}